\documentclass{aa}
\usepackage{natbib}
\usepackage{graphicx}
\usepackage{latexsym}
\newcommand{\nc}{\newcommand}  
\newcommand\nodata{ ~$\cdots$~ }% 
\nc{\teff}{$T_{\rm eff}$\,}  
\nc{\logg}{log\,$g$\,}  
\nc{\kms}{\,${\rm km\,s}^{-1}$\,}  
\nc{\mic}{$\xi_{\rm t}$\,}
\begin{document}

\title{{First stars IV.  
CS~29497-030: Evidence for operation of the $s$-process at very 
low metallicity}
\subtitle{}
\thanks {Based on observations made with the ESO Very Large Telescope 
at Paranal Observatory, Chile (program ID 165.N-0276(A)).}
}

\author{
T. Sivarani \inst{1} \and
P. Bonifacio\inst {1} \and
P. Molaro\inst {1} \and
R. Cayrel\inst {2} \and
M. Spite\inst {2} \and
F. Spite \inst {2} \and
B. Plez\inst {3} \and
J. Andersen\inst {4} \and
\hbox{B. Barbuy\inst {5}} \and
T. C. Beers\inst {6} \and
E. Depagne \inst{7} \and
V. Hill\inst {2} \and 
P. Fran\c cois  \inst{2} \and
B. Nordstr\"om \inst {8,4} \and 
F. Primas\inst {9}
} 

\institute{ 
    Istituto Nazionale di Astrofisica - Osservatorio Astronomico di
Trieste,
    Via Tiepolo 11, I-34131
             Trieste, Italy\\
   \email {sivarani@ts.astro.it,bonifaci@ts.astro.it,molaro@ts.astro.it}
         \and
              Observatoire de Paris-Meudon, GEPI,
             F-92195 Meudon Cedex, France\\
   \email {Vanessa.Hill@obspm.fr, Roger.Cayrel@obspm.fr}\\
   \email {Monique.Spite@obspm.fr, Francois.Spite@obspm.fr,
Patrick.Francois@obspm.fr }
         \and
             GRAAL, Universit\'e de Montpellier II, F-34095 
Montpellier
             Cedex 05, France\\
   \email {Bertrand.Plez@graal.univ-montp2.fr}
         \and
         Astronomical Observatory, NBIfAFG, Juliane Maries Vej 30,
         DK-2100 Copenhagen, Denmark\\
   \email {ja@astro.ku.dk}
         \and
    IAG, Universidade de S\~ao Paulo, Departamento de Astronomia, CP
         3386, 01060-970 S\~ao Paulo, Brazil\\
   \email {barbuy@astro.iag.usp.br}
               \and
             Department of Physics \& Astronomy, Michigan State 
University,
             East Lansing, MI 48824, USA\\
   \email {beers@pa.msu.edu}
\and
European Southern Observatory, Casilla 19001, Santiago, Chile\\
\email{edepagne@eso.org}
  \and
        Lund Observatory, Box 43, S-221 00 Lund, Sweden\\
   \email {birgitta@astro.lu.se}
        \and 
         European Southern Observatory (ESO),
         Karl-Schwarschild-Str. 2, D-85749 Garching b. M\"unchen, 
Germany\\
   \email {fprimas@eso.org}
}
\authorrunning{Sivarani et al.}
\titlerunning{First stars IV. CS~29497-030}
\offprints{T. Sivarani}
\date{Received xxx; Accepted xxx}
\abstract{

We present an abundance analysis of the very metal-poor, carbon-enhanced star
CS~29497-030. Our results indicate that this unusually hot turnoff star 
(T$_{eff} = 6650$K, log$g =$ 3.5) has a metallicity [Fe/H] $=
-2.8$, and exhibits large overabundances of carbon ([C/Fe] = +2.38), nitrogen
([N/Fe]= +1.88), and oxygen ([O/Fe]= +1.67). This star also exhibits a large
enhancement in its neutron-capture elements; the pattern follows that expected
to arise from the $s-$process. In particular, the Pb abundance is found to be
very high with respect to iron ([Pb/Fe]= +3.5), and also with respect to the
second peak $s$-process elements (e.g., Ba, La, Ce, Nd), which fits into the newly
introduced classification of lead (Pb) stars. The known spectroscopic binary
status of this star, along with the observed $s$-process abundance pattern,
suggest that it has accreted matter from a companion, which formerly was
an Asymptotic Giant-Branch
(AGB) star. In a preliminary analysis, we have also identified broad
absorption lines of metallic species that suggest a large axial rotational
velocity for this star, which may be the result of spin-up associated with the
accretion of material from its previous AGB companion. In addition, this star is
clearly depleted in the light element Li. When considered along with its rather
high inferred temperature, these observations are consistent
with the expected properties of a very low metallicity halo blue straggler.

}
\maketitle
\markboth{CS~29497-030}{}

\section{Introduction}

The two neutron-capture processes, the {\em slow} ($s-$ process), and the {\em
rapid} ($r-$process), occur under different physical conditions, and therefore
are likely to arise in different astrophysical sites. In an early systematic
study of elemental abundances in halo stars, \citet{spite} found that the
neutron-capture elements Ba and Y were over-deficient with respect to iron, and that
the barium abundance increased almost as fast as iron at low metallicity,
suggesting a common origin in massive, short-lived stars. From this
suggestion, \citet{truran} formulated the hypothesis that in the early Galaxy
the neutron-capture elements were formed exclusively through the $r$-process,
assumed to occur in core-collapse Type II SNe; this idea is often referred to as
the ``r-only hypothesis.'' More recent observations of heavy-element abundance 
patterns in metal-poor stars have further supported this hypothesis
\citep[~and references therein]{gilroy,burris}. 

From the theoretical point of view, support for the ``$r$-only hypothesis'' comes
from consideration of the timescales involved. The dominant site of the
$s$-process is thought to be the Asymptotic Giant-Branch (AGB) phase in
intermediate-mass stars \citep[ and references therein]{busso99}; by the time
the first of these objects began polluting the interstellar medium with
$s$-process elements, core-collapse SNe had already raised the mean metallicity
of the early-Galaxy Inter-Stellar Medium (ISM) above [Fe/H]$=-2.5$. Previous
stellar evolutionary models predicted that zero-metallicity stars would not undergo
the thermally pulsing (TP) AGB phase \citep{fuji}. Recently, however,
\cite{chieffi01} showed that thermal pulses can in fact occur even in stars
with no heavy elements. Although the environment might thus 
be appropriate, the
lack of iron seeds in such stars is still
thought to prevent the operation of the
$s$-process, hence no neutron-capture elements are predicted to be produced by
standard models of zero-metal TP-AGB stars \citep{abia,chieffi01,siess,iwamoto}. 

A growing body of observational evidence now suggests that, in contrast to the
$r$-only hypothesis, the $s$-process could indeed operate even at very low
metallicities. Whether or not it had a significant impact on the chemical evolution of
neutron-capture elements throughout the early Galaxy is still an open question,
as this depends on the Initial Mass Function of early-generation stars \citep{abia}. 

The large modern surveys for metal-poor stars
(the HK survey \citealt{beers85,
beers92, beers99,b99a}, and the Hamburg/ESO survey of \citealt{christ}) have
unveiled the existence of stars in which the abundances of neutron-capture
elements are greatly enhanced, although there are some important differences
between various classes of stars that exhibit this phenomenon,
which can be broadly grouped as described below.

A few stars, such as CS~22892-052 \citep{sneden94,sneden96,sneden00,sneden03},
and CS~31082-001 \citep{cayrel01,hill}, exhibit large enhancements of 
neutron-capture elements, including species that can only be synthesized
through the $r$-process, such as Th and U. The observed abundance 
patterns of the heavy
neutron-capture elements in these stars agree extremely well with a
scaled solar-system $r$-process pattern, suggesting an $r$-process origin
for all of them.  These highly $r$-process-enhanced metal-poor stars appear to be
extremely rare, with a frequency no greater than about 3\%
among giants with [Fe/H] $ < -2.5$. It should be
noted that although CS~22892-052 is enhanced in carbon ([C/Fe] $\sim$ +1.0;
\citealt{sneden94}), it has been suggested \citep{aoki02a} that the origin of
the carbon in this star may not be causally connected to the origin of
its $r$-process elements. CS~31082-001, on the other hand, is only mildly
carbon-enhanced ([C/Fe] = +0.2; \citealt{hill}).

Other very metal-poor stars in which the pattern of neutron-capture elements
suggest an $s$-process enrichment are more common. Radial velocity variations
have been observed in some \citep{norris97,PS2000,aoki02b}, but not all of them 
\citep[e.g., LP~706-7;][]{norris97}. In many of these
$s$-process-enriched, metal-poor stars the element lead is detected and 
often considerably enhanced. This is not unexpected, since the operation of the
$s$-process in a metal-poor star occurs with a high neutron-to-seed-nucleus
ratio, thus favouring the formation of neutron-rich nuclei.

Lead was first detected in a metal-poor star by \citet{cowan} in HD~126238, and
later in HD~115444 and CS~22892-052 \citep{sneden98,sneden00}. In our later
discussion, for CS~22892-052 we adopt the Pb abundance derived from the optical
lines, although it is very uncertain since the much stronger PbI 283.3nm
line has not been detected \citep{sneden03}. We prefer this
determination since the Pb abundance of CS~22892-052 estimated from the optical
lines is consistent with that obtained in the small number of other known
r-process-enhanced, metal-poor stars. In these three stars, Pb is attributed
primarily to the $r$-process rather than to the $s$-process because of the
large observed enhancements in other $r$-process elements such as Eu. 

Lead of likely $s$-process origin was first detected in the carbon- and
$s$-process-rich, metal-poor star LP~625-44 by \citet{aoki2000}, 
suggested by these authors to have originated from an AGB companion. 
Metal-poor
stars with high Pb abundances relative to the 
second  neutron-peak elements (Ba,
La, Ce), as predicted by metal-poor AGB models \citep{GM2000, GS01}, were
first discovered by \citet{vaneck}.  Subsequently,
Pb has been detected in 21 more stars \citep{aoki01,travaglio,aoki02b,johnson02,
cohen03,lucatello, vaneck03}. These stars show a large range in [hs/Pb] ratio 
(hs representing second $s$-process-peak elements such as Ba, La, Ce and
Nd). At present there appears to be  no evidence for a correlation between
[hs/Pb] ratio and metallicity of the lead stars. 

All the $s$-process-enhanced stars also appear to be enhanced in carbon,
hence one might speculate that they are analogs of the classical CH stars, all
of which appear to be members of binary systems 
\citep{McClure1990,McClure1997}. If this is
universally true, the $s$-process elements and carbon have almost certainly been
produced by a binary companion during its TP-AGB phase, and subsequently
accreted onto the surviving companion star. To complete this (already complex)
picture one must add that the reverse situation is not true; not all of the
carbon-rich stars, which appear to represent as many as 15-25\% of all stars
with [Fe/H] $< -2.5$ \citep{beers85,beers92,beers99,b99a, christ}, are also
enhanced in their neutron-capture elements. In fact, there exist
carbon-enhanced, metal-poor stars with no observed enhancements of
neutron-capture elements at all (e.g., CS~22957-027,
\citealt{norris97,boni98}). Recently, \citet{aoki02b} have enlarged the number of
such stars, suggesting that the phenomenon of mild carbon-enhancement without
accompanying neutron-capture enhancement amongst very metal-poor stars is not
uncommon. In addition, there exists a class of extremely metal-poor stars
(including CS~22949-037, \citealt{depagne}, and CS~29498-043, \citealt{aoki02c})
that exhibit large excesses of N, O, Mg, and Si, in addition to C, with no
apparent neutron-capture enhancement. It is thus not clear what relation, if any, 
exists between the carbon and neutron-capture-element enhancement phenomena.
 
\begin{figure}
\rotatebox{90}{\resizebox{7.5cm}{!}{\includegraphics{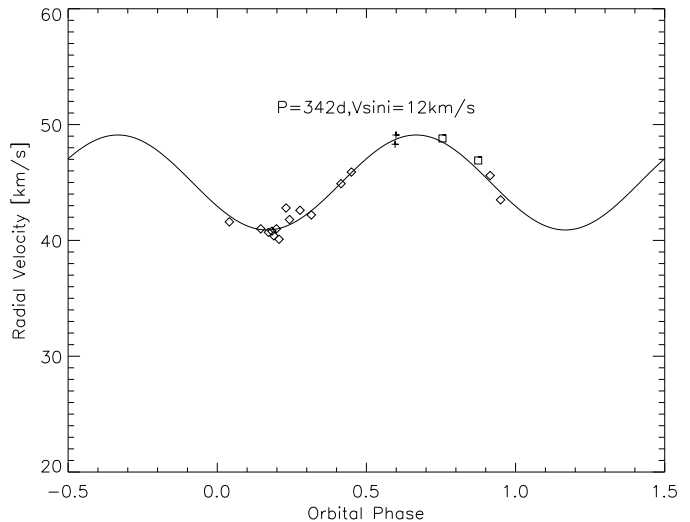}}}
\caption{The spectroscopic orbit of CS~29497-030 by \citet{sneden03}; the data 
by \citet{PS2000} are shown by ``$\Diamond$'', our data by ''+'' symbols. }
\label{RV}
\end{figure}

In this paper we report a detailed high-resolution analysis of CS~29497-030,
based on high-quality VLT/UVES spectra, which we show to be a very metal-poor
star that exhibits enhanced abundances of neutron-capture elements with a
pattern suggesting an $s-$process origin. CS~29497-030 was originally
classified  as A type by G. Preston from
inspection of its objective prism image. 
The star was included as BHB candidate in the
UBV survey of \citet{preston91a} but subsequently 
rejected as a BHB star
on the basis of the two colour criterion
by \citet{preston91b}.
It was subsequently
identified as a high-gravity metal-deficient (BMP) star on the basis of its
small B-V and U-B excess by \citet{PBS94}.
As such it was also included in the spectroscopic survey  of
\citet{Wilhelm}.
In preparing our VLT observations T.C. Beers visually inspected
the medium survey spectra of all the stars considered for possible
follow-up at high resolution, and noticed a 
G-band  of  unusual strength compared to its
hydrogen lines, which are typical of stars at or
beyond  the main-sequence TO, and thus marked it as
a possibly carbon-rich star.
\citet{PS2000} included this star are part of
a long-term radial velocity survey of BMP stars, and identified CS 29497-030 as
a spectroscopic binary with a period of 342 days; the radial velocity curve is
shown in Fig. \ref{RV}. 

As briefly reported earlier by our group \citep{sivarani02}, the most striking
features of CS~29497-030, besides its very high C abundance, are its extremely
high Pb abundance, which places CS~29497-30 amongst the most metal-poor lead
stars yet identified, as well as its extremely high O abundance. CS~29497-30 is
thus only slightly less enhanced in O than the extreme case of CS~22949-037, an 
extremely metal-poor giant with no evidence of binarity \citep{depagne}.

\section{Observations and data reduction}

Our spectroscopic data were obtained at the VLT-Kuyen 8.2m telescope, using the UVES 
spectrograph \citep{dekker} at 
a resolving power of $R = 43000$. The observations
were made as a part of the Large Programme 
``First Stars''(165.N-0276; P.I. R. Cayrel); the log
of observations is given in Table 1. The data were reduced using the UVES
context within MIDAS, which includes bias subtraction, flat fielding, wavelength
calibration, and merging of echelle orders. The continuum normalisation was
performed with IRAF (using a cubic spline function) for the merged spectra. For
a few lines that were either very weak or of particular interest, we have used
the spectra of the individual orders without merging them. The Balmer lines of
hydrogen were checked with both single-order and merged spectra.
Equivalent-width measurements for unblended lines were accomplished by fitting
gaussian profiles, using the genetic algorithm code described in
\citet{francois}. 

Radial velocities were measured from the positions of unblended lines in the
range 480-850nm. Small instrumental shifts were corrected using the telluric
absorption lines. As can be seen from inspection of Figure 1, the new radial
velocity data for CS~29497-030 provide tight constraints on the binary orbit, as
they were obtained near one of the peaks in the velocity curve. The
orbit calculated by \citet{PS2000} is clearly confirmed by our new data .

\begin{table}
\label{log}
\caption{Log of observations}
\begin{tabular}{llllll}
\hline
Date        & $t_{exp}$(s) & $\lambda_c$ (nm) & S/N$^{*}$ & MJD-24000.5& $v_R$ (\kms )  \\

\hline
19/10/2000  & 1800         & 396+850          & 75 & 51836.20 &  48.8 \\
06/11/2001  & 1800         & 396+573          & 65 & 52219.17 &  46.9 \\
\hline
\end{tabular}

$^{*}$ S/N at 440nm
\end{table}

\section{Stellar parameters and analysis}

CS~29497-030 was classified as an A~V star by \citet{Wilhelm}, indicating that
it was thought to have a high surface gravity, at least as compared to the field
horizontal-branch stars they were seeking to identify. They determined
the stellar parameters given in Table \ref{literature}, based on available $UBV$
photometry and medium-resolution ($\sim 1.5$ \AA)  spectroscopy. 
Recently, \citet{KC2002} have derived temperatures based on
the $V-J$, $V-H$, and $V-K$ colours, using 2MASS data and fitting the colours to
Kurucz model predictions. 

We have obtained new photometric data, $BVR_{C}I_{C}$ (the subscript ``C'' indicates
the Cousins system) with the Danish 1.5m telescope and DFOSC instrument
\citep{beers03}. We estimate a color excess due to reddening of $E(B-V)$ $=$
0.016 from the maps of \citet{Sch98} and use the
\citet{alonso} calibrations to derive \teff from the colours. The Alonso
calibrations of $B-V$, $V-R$, $V-I$, $R-I$, and $V-K$ require colours in the
Johnson system; calibrations of the IR colours, $J-H$ and $J-K$, are in the 
TCS
system (the native photometric system at the 1.54m Carlos S\'anchez telescope 
in Tenerife; \citealt{arribas}).

To transform the colours from the Cousins to the Johnson system, we use the
transformations derived by \citet{bessell83}, which for $R-I$ do not include the
Paschen dip; \citeauthor{bessell83} points out that, as a result, these
transformations are uncertain for $-0.05 \le R-I \le +0.3$. CS~29497-030 falls
in this colour range, where the Paschen jump reaches a maximum. To obtain the $V-K$
colour in the Johnson system we transformed the 2MASS $K$ magnitude to the
\citet{BesselBrett} homogenized system, using equations from Explanatory Supplement
 to the 2MASS All Sky Data Release \citep{cutri}; note that this is a two-step calibration, 
2MASS to CIT (the CIT system is described in 
\citealt{elias} and \citealt{frogel}), and then to the  \citeauthor{BesselBrett}
 system. Therefore, the  errors involved in the colour transformations must be 
added to those of the photometry. 

To transform the 2MASS colours $J-H$ and $J-K$ onto the TCS system 
\citep{alonso}
again
required a two-step
calibration, from 2MASS to CIT \citep{cutri}, and then from CIT
to TCS \citep{alonso}. 
In Table \ref{atmos}, we list the errors in \teff corresponding
to the photometric errors. 

The temperatures listed in Table \ref{atmos} span a range of almost 1000 K. The
$B-V$, $J-H$ and $J-K$ colours provide similar temperatures, while the $V-R$,
$V-I$, and $V-K$ colours predict higher temperatures. The large errors in the
temperatures derived from $J-H$ and $J-K$ are due to the very steep slope of the
calibrations for the bluest stars. With this large spread in temperatures there
is little point in averaging them. We therefore decided to rely on the Fe I
excitation equilibrium, which implies \teff = 6650K, similar to the \teff
derived from the $B-V$, $J-H$, and $J-K$ colours. The wings of the Balmer lines
are also consistent with this temperature. The presence of a CN molecular band
in the spectra suggests that the temperatures cannot be as high as derived from
the $R-I$, $V-I$, and $V-K$  colours. The only explanation  we can offer
for the large spread in effective temperatures derived from different colours is
that the star has an anomalous spectral energy distribution, possibly due to
its peculiar composition and/or its binary nature.

The FeI/FeII ionisation equilibrium has been used to fix the surface gravity of
CS~29497-030. We point out that, with our adopted log g, balance is achieved,
within the errors, also for TiI/TiII, MgI/MgII and MnI/MnII,
although TiI and MgII are represented only by single lines. 
Dr. A. Korn kindly
performed NLTE computations for the iron lines in order to check for NLTE
effects on the derived gravity; he obtained log g(NLTE) = 3.65, confirming
a rather 
low surface gravity for this star. 
Also, the observed $U-B$ colour of the star is
compatible with this gravity.

The microturbulence was determined in the usual way, 
by requiring that the
abundances derived from the Fe I lines be independent of the measured equivalent widths.

The model atmospheres (OSMARCS), spectrum synthesis code (TURBOSPEC by Plez), and
the line data that we employed are the same as described in previous papers of this series
\citep{hill,depagne,francois}.

\begin{table}
\caption{Photometry of CS~29497$-$030} 
\label{atmos}
\begin{tabular}{llll}
\hline
Mag \& colour    &                     & \teff         &$\delta$\teff$^{a}$\\
\hline
$V$                  & 12.65 $\pm 0.002$   & \nodata       & \nodata \\
$U-B^{b}$            & $-$0.14             & \nodata       & \nodata \\
$B-V$                & 0.299 $\pm 0.004$   & 6680          & 20      \\
$V-R_{C}$            & 0.215 $\pm 0.003$   & 6875          & 35      \\
$V-I_{C}$            & 0.440 $\pm 0.005$   & 7016          & 25      \\
$R-I_{C}$            & 0.225 $\pm 0.008$   & 7281          & 104     \\
$V-K_{2Mass}$        & 0.905 $\pm 0.025$   & 7153          & 59      \\
$J-H_{2Mass}$        & 0.183 $\pm 0.038$   & 6693          & 303     \\
$J-K_{2Mass}$        & 0.217 $\pm 0.036$   & 6563          & 257     \\
\hline
\end{tabular} 
\noindent \parbox[t]{8cm}{
{\it $^{a}$ Change in \teff for a $\pm$1$\sigma$ change in the colour.}\\
{\it $^{b}$ \citet{PS2000} }\\}
\end{table}

\begin{table}
\caption{Atmospheric Parameters of CS 29497$-$030} 
\label{literature}
\begin{tabular}{lllll}
\hline
Reference         & \teff & log g(cgs) & [Fe/H] & \mic \kms \\
\hline
 1   & 7426  &  3.9  &  $-$2.5  & \nodata  \\
 2   & 7180  &  4.2  &  $-$2.0  & \nodata  \\
 3   & 7050  &  4.2  &  $-$2.16 & 1.75    \\
 4   & 6650  &  3.5  &  $-$2.7  & 2.0  \\
\hline
\end{tabular}
\noindent \parbox[t]{8cm}{
(1) \citet{Wilhelm},  (2) \citet{KC2002}, 
(3) \citet{sneden03} (4) Present Work}

\end{table}

\section{Abundances}

\begin{figure}
\rotatebox{90}{\resizebox{7.5cm}{!}{\includegraphics{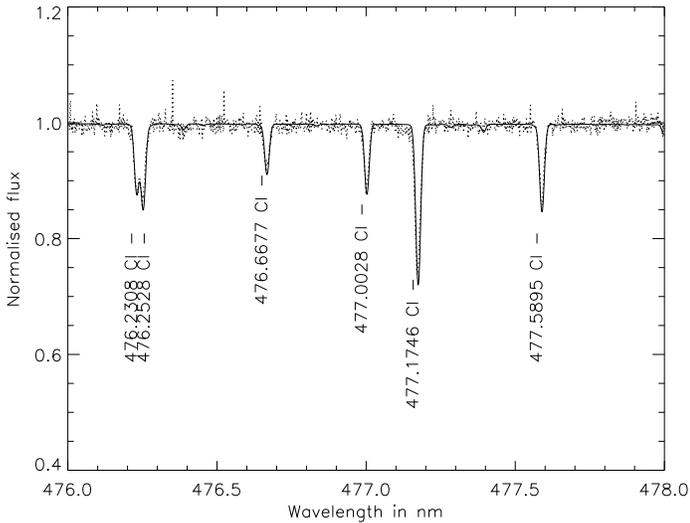}}}
\caption{Examples of the high-excitation
CI lines observed in CS~29497-030. The solid line
is the synthetic spectrum (for [O/Fe] = +1.67) computed
with the fitted C abundance, which is 0.3 dex higher than what is derived from
the CH lines. A probable cause of this discrepancy is NLTE effects, such as
those that affect the OI 777nm triplet.}
\label{CI}
\end{figure}

\begin{figure}
\rotatebox{90}{\resizebox{7.5cm}{!}{\includegraphics{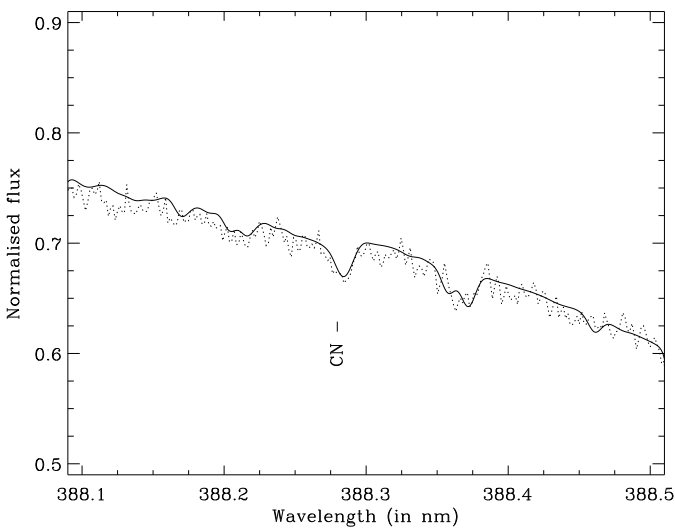}}}
\caption{The CN 388nm bandhead.  The dotted line represents the observed
spectrum, while the solid line is the synthetic spectrum (for [O/Fe] = +1.67).} 
\label{cn}
\end{figure}

\begin{figure}
\rotatebox{90}{\resizebox{7.5cm}{!}{\includegraphics{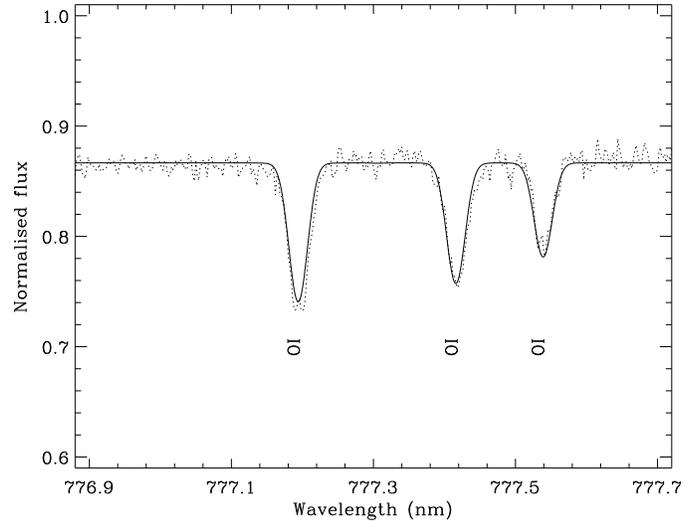}}}
\caption{The OI 777nm triplet.  In this and the following figures, the dotted line 
represents the observed spectrum, while the solid line is the synthetic spectrum.}
\label{OI}
\end{figure}

\begin{figure}
\rotatebox{90}{\resizebox{7.5cm}{!}{\includegraphics{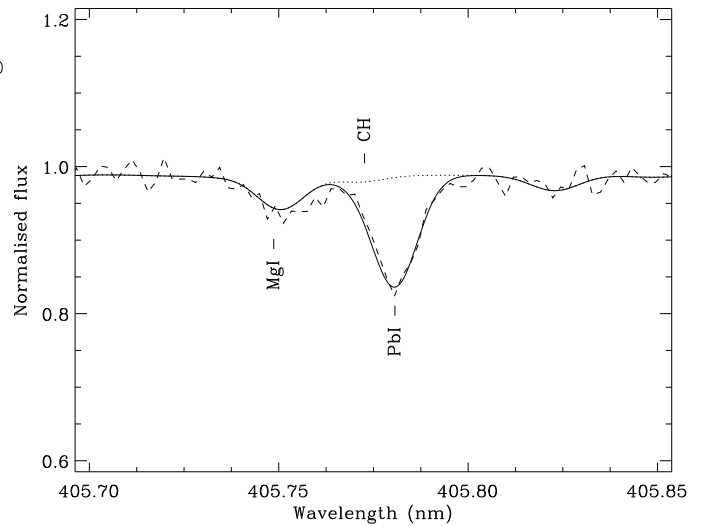}}}
\caption{The Pb I 405.7nm line.
The dotted line shows the (small) contribution of the CH
line blend (for [O/Fe]= +1.67). }

\label{Pb}
\end{figure}

\begin{figure}
\rotatebox{90}{\resizebox{7.5cm}{!}{\includegraphics{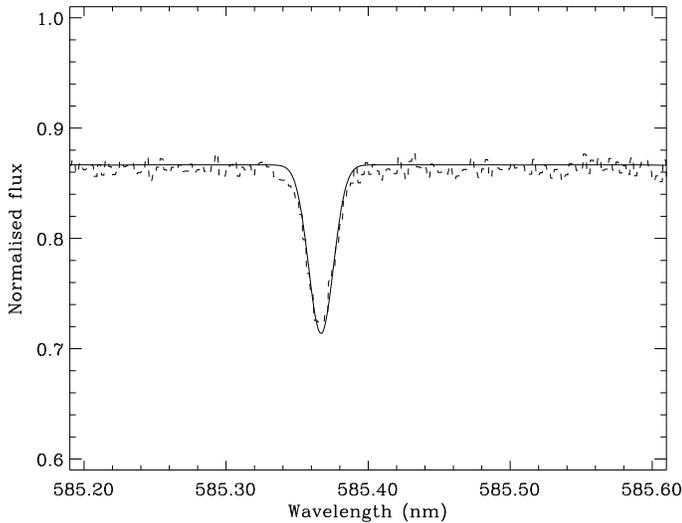}}}
\caption{The  Ba II 585nm line. }
\label{Ba}
\end{figure}

\begin{figure}
\centering
\rotatebox{90}{\resizebox{7.5cm}{!}{\includegraphics{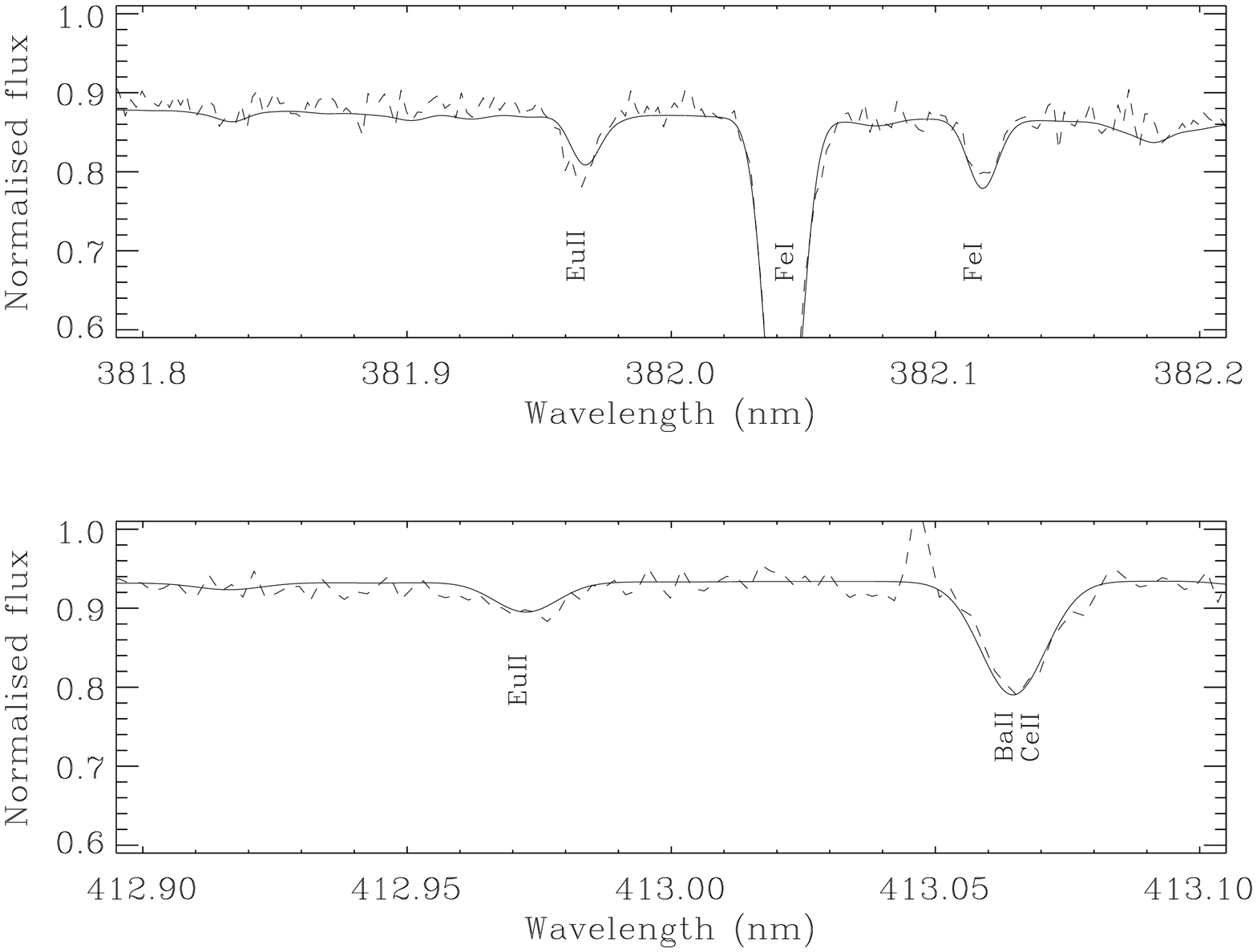}}}
\caption{Two Eu II lines.  }
\label{Eu}
\end{figure}

The derived elemental abundances for CS~29497-030 are listed in Table
\ref{abun}. The most striking features are the large overabundances of C, N, and
O, relative to iron, and the large enhancement of all the $s$-process elements, among
which Pb stands out with [Pb/Fe] = +3.5.
 
The chemical composition of CS~29497-030 was analysed earlier by
\citet{PS2000}, who derived abundances for Fe, Mg, Ca, Sc, Ti, Cr, Mn, Sr, and Ba,
and by \citet{sneden03b}, who derived in addition abundances for C, O and La, Nd, Eu and
Pb.
The overall abundance pattern derived by both
the above  analyses is similar to our own, 
with most elemental ratios showing agreement at a level of 0.2--0.3 dex. 
However,  both obtain a
significantly higher metallicity ([Fe/H] = --2.1, --2.16, respectively) 
 than we do, probably due to the 
higher temperature adopted.

\subsection{Carbon and Nitrogen}

We measured several CI and CH lines in our spectra. The CI lines, some of which 
are shown in Fig. \ref{CI},
provide a C abundance that is about 0.3 dex higher than that derived from the CH
lines. We suspect the difference could be due to NLTE effects in the CI lines,
which have excitation potentials higher than 7.5 eV. We plan to perform NLTE
computations in the near future to check on this possibility. Taking into
account all the CI lines in the spectra, we obtain an abundance which is on
average 0.1--0.3 dex higher than obtained from the CH lines. The $^{13}$CH lines are very
weak, so we could only derive a lower limit for the carbon isotope ratio:
$^{12}$C/$^{13}$C $>$ 10.

We have detected a very weak CN 388nm bandhead, shown in Fig. \ref{cn}, from
which, assuming the C abundance derived from the CH lines, we derive [N/Fe] =
+1.88. 

We have checked the sensitivity of the derived carbon abundance to the adopted
oxygen abundance by performing the analysis both with the high oxygen abundance 
derived from the OI 777.4nm lines ([O/Fe] = +1.67), and for a much lower value 
([O/Fe] = +0.5). We find no impact of these two widely different oxygen abundances 
on the derived carbon and nitrogen abundances.

\subsection{The $\alpha$ elements}

The oxygen abundance was derived from the OI 777.4nm triplet shown in Fig.
\ref{OI}. The oxygen triplet is known to suffer from NLTE effects \citep{kiselman91,
takeda,gratton,kiselman01}. In order to correct for these effects we
interpolated in Table 10 of \citet{gratton}  and found a correction of --0.09
dex; the NLTE corrected value is listed in Table \ref{abun}. This oxygen
abundance appears to be exceptionally high when compared to most other
metal--poor stars \citep{depagne}. Note that the high abundance derived from the
OI 777.4nm triplet is still consistent with the non-detection of the 615nm OI
triplet and 630nm [OI] lines. There is always some concern in deriving
abundances from the OI triplet. For example, for two giants in the LMC,
\citet{barbuy94} report an overabundance of about 2.0 dex compared to the
abundance derived from the 615.8nm lines. On the other hand, there is at least
one star, the very metal-poor subgiant BD +23$^{o}$ 3130, for which the oxygen
abundance derived from the OI 777.4nm triplet does agree with that derived from the
[OI] 630nm line and the OH UV lines \citep{cayrel01a,garcia01}. In the
comprehensive study of
\citet{carretta}, oxygen abundances were derived both from permitted OI lines
(the 616nm triplet and the 777nm triplet) and from the [OI] 630nm line.  There appears
to be no systematic difference between these sets of abundances, as illustrated
by their figure 3. In the case of CS~29497-030, the OI triplet is the only available oxygen
indicator, hence our only choice is to take the derived oxygen abundance at face
value, although with some caution.

Magnesium exhibits a ``normal'' overabundance of $\sim 0.5$ dex above Fe. The
MgII 448.1nm doublet is detected, testifying to the relatively high effective
temperature of this star. The derived abundance from the MgII doublet is only
0.1 dex higher than that derived from the MgI lines. Considering that the latter
show a scatter of 0.2 dex, we consider the two abundances to be consistent. 

We have detected a strong Si I line at 390.5nm, and weak 
Si II lines at 385.6nm
and 386.2nm; the derived abundances are [SiI/Fe]=--0.05 and [SiII/Fe]= +0.6,
respectively. We have spectra taken on two different nights for this wavelength
region; we find they are consistent with one another; therefore the problem 
should not 
originate in the data. The Si I line is saturated and sensitive to the 
adopted microturbulence, while the Si II lines are sensitive to log g. By 
adjusting the atmospheric parameters we obtain identical abundances 
([Si/Fe]= +0.4) for Si I and Si II for a model with \teff = 6750K,
log g = 3.0, 
\mic = 0.0\kms. It is interesting to note that \citet{McWil95} find
a large dispersion in the Si abundance in a sample of 33 very metal-poor stars,
based primarily on abundances estimated from the Si I 390.5nm line. 
Also, there is a
large difference seen between their Mg and Si abundances. We adopt the Si
abundance from the Si II lines, because it is consistent with the level of
overabundance seen in Mg. 

The sulphur abundance is very uncertain because it relies on a single line at
912.28nm, which is contaminated by a telluric absorption feature; we estimate an
error of 0.3 dex. Assuming that our process of dividing out the telluric feature
has been successful, sulphur is found to be overabundant by about 0.4 dex with
respect to iron, in line with Mg. The general 
behaviour of S in metal-poor stars
is that sulphur is enhanced over iron, [S/Fe]$>0$ \citep{fra1,fra2,ir01,chen}.
The data at very low metallicity is still rather limited, hence we cannot
comment here on any general trend with declining metallicity.

Ca and Ti exhibit ``normal'' overabundances with respect to iron, of the order
of 0.3 dex.

\subsection{The odd-Z elements}

The sodium abundance was obtained from the NaI D resonance lines, and displays a
marked overabundance with respect to iron of $\sim 0.5$ dex. The aluminium
abundance was derived from the resonance line at 394.4nm, and Al is  
underabundant by $\sim -0.7$ dex with respect to iron. 

These abundances are the LTE values; no NLTE corrections, as prescribed by 
\citet{Bau97} and \citet{Bau98}, have been applied.
The prescribed NLTE corrections for \teff = 6500K, log g = 4.0, [Fe/H] = --3.0
are --0.11 and +0.65 for Na and Al, respectively. If we assume these values to
apply for our lower-gravity star, then Al follows iron while Na
will still be overabundant by $\sim 0.4$ dex with respect to iron, reminiscent 
of the similar, but vastly larger overabundance of Na in CS~22949-037 \citep{depagne}.

\subsection{The iron-peak elements}

Our derived Fe abundance places CS~29497-030 among the very metal-poor stars,
with [Fe/H] $< -2.0$. Our Fe abundance and surface gravity were independently
checked by Andreas Korn, who found [Fe/H]= --2.71, log g = 3.40 in LTE, using a
separate set of atmospheric models and line data, and [Fe/H]= --2.59, log g =
3.65 in NLTE. 

Mn and Cr show a slight underabundance of $\sim 0.2$ dex compared to iron. This
is in line with the general decrease of the ratios of these elements relative to
iron initially found by \citet{McWil95} and confirmed by other authors
\citep{ryan96, carretta02,francois}. Cobalt, on the other hand, shows a slight
overabundance with respect to iron. This 
behaviour has also been seen before in
very metal-poor stars \citep{McWil95,ryan96,francois}.

\subsection{The neutron-capture elements}

All the $s$-process elements in CS~29497-030 exhibit overabundances with 
respect to iron, and the heavy $s$-process elements (hs: Ba, Ce, La, Nd) are more 
overabundant than the lighter $s$-process elements (ls: Sr, Y, Zr).
The Pb abundance was derived from the 405.7nm line shown in Fig. \ref{Pb}. The 
line data has been taken from \citet{vaneck03}; we have included the hyperfine 
splitting and isotopic shifts as given in \citet{vaneck03}. We also detect the 
PbI 368.3nm and 406.2nm lines in our spectra, from which we derive abundances 
that are consistent with the 405.7nm line. The derived abundance of Pb is very 
high, about 3.5 dex relative to iron, and also very high compared to those of 
Ba, La, Ce, and Nd (e.g., [Pb/Ba] = +1.38). 

The Sr abundance was determined from the Sr II resonance lines,
which are quite strong (equivalent widths of the order of 10 pm 
(where 1pm $= 10^{-12}$m). 
We detected seven BaII lines, including the resonance lines at 455.40nm and
493.40nm. In our analysis of Ba, we have adopted the hyperfine splitting
provided by \citet{McWil95}. We checked the effect of adopting either the
solar-system total abundance or the solar-system $s$-process isotopic
fraction, and found that the derived abundance does not change. For our final
determination we used only the weaker lines, like the 585nm line shown in Fig.
\ref{Ba}, since the stronger lines are clearly affected by saturation. Note that
increasing the microturbulence by only 0.5 \kms achieves a concordance between
Ba abundances derived from weak and strong lines. We thus believe our Ba
abundance to be very robust. 

La, Ce, and Nd abundances were determined using
spectrum synthesis. We have not used any hyperfine splitting for the La, Ce, Nd,
and Y lines, since they are weak and the abundances are not expected to be
affected by hyperfine splitting \citep{McWil95}. 
A total of six Eu II lines are detected in our spectra; two such lines are shown in
Figure \ref{Eu}. We have checked, using the hyperfine splitting provided by
\citet{kurucz}, that the derived abundance is insensitive to the use of
hyperfine splitting. The abundance listed in Table \ref{abun} has been obtained
using the $gf$ values from \citet{lawler}, and without hyperfine splitting.

\begin{table}
\caption{Elemental abundances for CS~29497-030}
\begin{center}
\begin{tabular}{llrrrr}
\hline
Species &A(X)$_\odot$ & [X/H]  & [X/Fe] & $\sigma$ & n       \\
\hline
Li I    & 1.10      & $<$ 0.00  & $<$2.7     & \nodata & \nodata \\   
C I     & 8.52      & $-$0.01    &    2.69    &  0.24   &  22     \\
CH      & \nodata   & $-$0.32    &    2.38    & \nodata & \nodata \\
CN      & 7.92      & $-$0.82    &    1.88    & \nodata & \nodata \\
O I     &  8.74     & $-$1.03    &    1.67    &  0.1    &  3      \\
NaI     & 6.33      & $-$2.18    &    0.52    &  0.15   &  2      \\
MgI     & 7.58      & $-$2.16    &    0.54    &  0.21   &  4      \\
MgII    & \nodata   & $-$2.06    &    0.64    & \nodata &  1      \\
Al I    &  6.47     & $-$3.37    &   $-$0.67  & \nodata &  1      \\
Si I    &  7.55     & $-$2.75    &   $-$0.05  & \nodata &  1      \\
Si II   &  7.55     & $-$2.10    &    0.60    & \nodata &  2      \\
S I     &  7.33     & $-$2.30    &    0.40    & \nodata &  1      \\
K I     &  5.12     &$< -$2.00  & $<$0.7     & \nodata & \nodata \\
Ca I    &  6.36     & $-$2.37    &    0.33    & 0.17    &  9      \\
Sc II   &  3.17     & $-$2.40    &    0.30    & 0.05    &  3      \\
Ti I    &  5.02     & $-$2.54    &    0.16    & \nodata &  1      \\
Ti II   & \nodata   & $-$2.41    &    0.29    & 0.28    &  19     \\
Cr I    &  5.67     & $-$2.87    &   $-$0.17  & 0.12    &  6      \\
Cr II   & \nodata   & $-$2.75    &   $-$0.05  & 0.07    &  4      \\
Mn I    &  5.39     & $-$2.89    &   $-$0.19  & 0.28    &  3      \\
Mn II   & \nodata   & $-$2.87    &   $-$0.17  & 0.00    &  2      \\
Fe I    &  7.50     & $-$2.77    &   $-$0.07  & 0.14    &  55     \\
Fe II   & \nodata   & $-$2.70    &   $-$0.00  & 0.10    &  5      \\
Co I    &  4.92     & $-$2.28    &    0.42    & 0.10    &  2      \\
Ni I    &  6.25     & $-$2.91    &   $-$0.21  & 0.22    &  11     \\
Zn I    &  4.60     &$< -$2.80  & $<$-0.10   & \nodata & \nodata \\
Sr II   &  2.97     & $-$1.86    &    0.84    & 0.05    &  2      \\
Y II    &  2.24     & $-$1.99    &    0.71    & 0.07    &  5  \\
Zr II   &  2.60     & $-$1.27    &    1.43    & 0.15    &  3      \\
Ba II   &  2.13     & $-$0.53    &    2.17    & 0.25    &  6      \\
La II   &  1.17     & $-$0.60    &    2.10    & 0.10    &  11     \\
Ce II   &  1.58     & $-$0.56    &    2.14    & 0.17    &  4 \\
Nd II   &  1.50     & $-$0.85    &    1.85    & 0.10    &  8      \\
Eu II   &  0.51     & $-$1.26    &    1.44    & 0.15    &  6      \\
Pb I    &  1.95     &    0.85   &    3.55    & \nodata &  1      \\
$^{12}$C/$^{13}$C     &&& $>$ 10 & \\
\hline
\end{tabular}
\end{center}
\label{abun}
\end{table}
\section{Discussion}

\subsection{$s-$process nucleosynthesis in metal-poor stars}

The most likely site for the the $s-$process is the inter-shell region of a
thermally pulsating AGB star, provided a suitable neutron source is active.
Until recently it was believed that stars of zero or extremely low metallicity
did not experience the TP-AGB phase \citep{fuji}. In contrast, \citet{chieffi01} 
have recently shown that zero-metallicity stars {\it do} undergo thermal pulses, 
although by a somewhat different mechanism, at least for stars in the range
$4M_\odot\le M\le 6M_\odot$. These results have been independently confirmed by
\citet{siess}. 

Our present understanding of the behaviour of $s-$process nucleosynthesis has
been reviewed by \citet{busso99}. There is a general consensus that the neutron
source is the reaction $\rm ^{13}C(\alpha,n)^{16}O$. In order to activate it, a
partial mixing of protons into the C-rich layer is required. The hydrogen is
burned through $\rm ^{12}C(p,\gamma)^{13}N(e^+\nu) ^{13}C(p,\gamma)^{14}N$,
leaving a residual $^{13}\rm C$ abundance \citep{ir}. Although the majority of
workers in the field admit that this partial mixing does take place, there is no
consensus on the {\it amount} of mixing, nor on the profile of the resulting
$^{13}$C {\em pocket}. The usual practice has been to simply assume the amount
of $^{13}$C \citep{gall98, GM2000,busso01,GS01}. 

Zero-metal stars are a special case in that one could expect the lack of Fe seeds  to 
prevent the operation of the $s-$process even in the presence of a suitable
neutron flux. However, \citet{GS01} showed that if there is a partial mixing of
protons, then the $s-$process proceeds, starting from $\rm ^{12}C$, and builds
all the heavy metals up to Pb and Bi. All of these authors agree on
the existence of a large enhancement of lead with respect to the nearby
second-peak $s-$process elements (Ba, La, Ce, Nd). This is a consequence of the
large neutron-to-seed ratio.

Recently, \citet{iwamoto} have proposed a new $s$-process paradigm, in which the
abundances of the neutron-capture elements result from only one (or at most a few) 
neutron exposures. They find that, in models of metallicity less than [Fe/H]=
--2.5, the helium convective shell penetrates the H-rich layers, allowing
proton mixing to occur. In their scenario, they achieve a high neutron flux, and
almost all of the $s-$process elements (except Pb) are made in the first
neutron irradiation. The abundance of Pb is more sensitive to the number of
pulses. From their models they were able to explain the observed abundance patterns
of LP~625-44 and LP~706-7, which have Pb/Ba $\sim$+1.0 and could not be
easily explained by standard partial-mixing scenarios \citep{gall98,GM2000,
busso01,GS01}.

\subsection{The evolutionary status of  CS~29497-030 }

Compared to most other very metal--poor stars, CS~29497-030 stands out for its 
high effective temperature. There is little doubt that the star is indeed hotter 
than most other metal--poor stars that have been analysed to date, because we 
detect the Mg II 448nm doublet. This immediately poses an age problem, because 
the star is considerably hotter than the main-sequence turnoff (TO) of theoretical 
isochrones of age 10 Gyr and comparable metallicity
\citep[$Z=0.0001$;~][]{Gir02}. Our derived gravity also implies that the star is
more luminous than the TO, however it is not sufficiently luminous to be
assigned to the horizontal-branch stage of evolution.
Its evolutionary status thus appears to be a puzzle. 

It is
likely that the anomalous position of CS~29497-030 in the (log T, log g ) plane
is due to its binary nature and past accretion history. Given that the Halo TO
is considerably redder than this star, it is legitimate to call it a ``Halo Blue
Straggler.'' It was in fact expected that some of the BMP stars (which this
object has been identified as) are indeed bona-fide Blue Stragglers.
 It is interesting to notice  that one
of the principal conclusions of
the radial velocity
investigation of \citet{PS2000} was that  many candidate BMP stars
are indeed halo binary blue stragglers formed by mass
transfer rather than merger.

We note that the LiI 670.8nm doublet is not detected in our spectra. This is
consistent with the identification of CS~29497-030 as a blue straggler, since
such stars are observed to be highly Li-depleted \citep{glaspey}. 
\citet{ryan02} identified several ultra Li-depleted halo stars that exhibit
large projected 
rotational  velocities, which prompted them to suggest that these
stars had experienced previous transfer of mass and 
angular momentum from a massive companion.
\citet{PS2000} obtained a rather large rotational velocity ($v\sin i = 12$\kms)
for CS~29497-030.  
All our spectra, at both observed epochs, show lines with a FWHM
of $\sim 9.55$ \kms, which is larger than the $\sim 7 $ \kms expected
from our instrumental resolution. If interpreted as due to rotation,
the excess broadening would be of the order of  $\sim$ 6.5 \kms.
Such rotational velocities
are unexpected for old halo stars and hence provide evidence that
the surviving star of the binary 
system has been significantly spun up due to
mass transfer. In fact this is far larger than that
corresponding to synchronous rotation of the star in its
342-day circular orbi, which would be below 0.5 \kms.

\begin{figure}
\resizebox{\hsize}{!}{\includegraphics{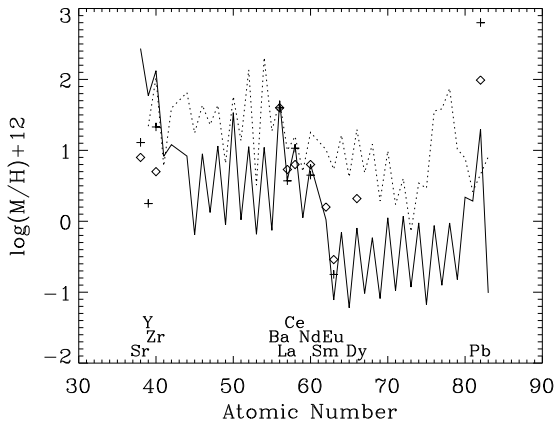}}
\caption{Abundance patterns of $s$-process enhanced stars.  The solid line
indicates the main solar-system $s$-process component determined by
\citet{Arlandini}, while the dotted line indicates the $r$-process component.
The ``+'' symbol indicates the abundances for CS~29497-030; the ``$\diamond$''
represents CS~31062-050 (from \citealt{aoki02b}. The abundance patterns are normalised
to Ba.}
\label{spatt_tab}
\end{figure}

\begin{figure}
\resizebox{\hsize}{!}{\includegraphics{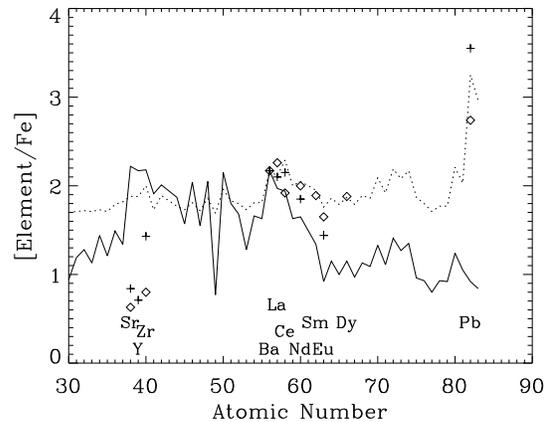}}
\caption{Abundance patterns of $s$-process enhanced stars. The solid line indicates
the surface abundance of an AGB model with solar
metallicity, while the dotted line indicates the surface abundance of an AGB
model with Z=0.001 after 50 thermal pulses \citep{GM2000}. The ``+'' symbol indicates the
abundances for CS~29497-030; the ``$\diamond$'' represents CS~31062-050 (from
\citet{aoki02b}. The abundance patterns are normalised to Ba.}
\label{spatt_tab1}
\end{figure}

\begin{figure}
\resizebox{\hsize}{!}{\includegraphics{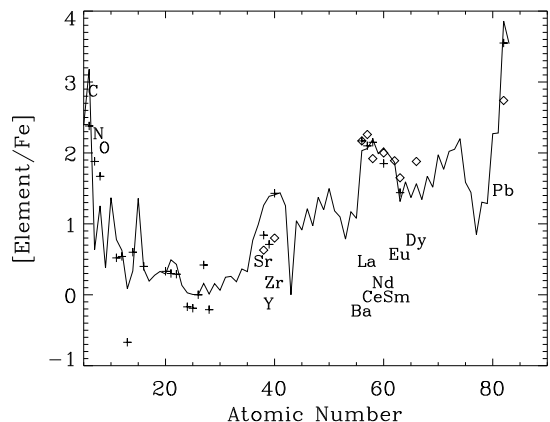}}
\caption{
Comparison between the observed abundances in
CS~29497-030 ( ``+'' symbol ) and  CS~31062-050, from
\citet{aoki02b}
(``$\diamond$''), with the theoretical $s-$process 
nucleosynthesis computations
of Gallino et al. (in preparation)  for accretion
from a 1.3 M$\odot$ star in the AGB phase through
stellar wind. The abundances from the AGB star are
considered to be diluted by a factor of 2 by the
unprocessed material of the envelope of the companion star.
No normalization of abundances has been applied.
}
\label{gallino}
\end{figure}

\begin{figure}
\rotatebox{90}{\resizebox{7.5cm}{!}{\includegraphics{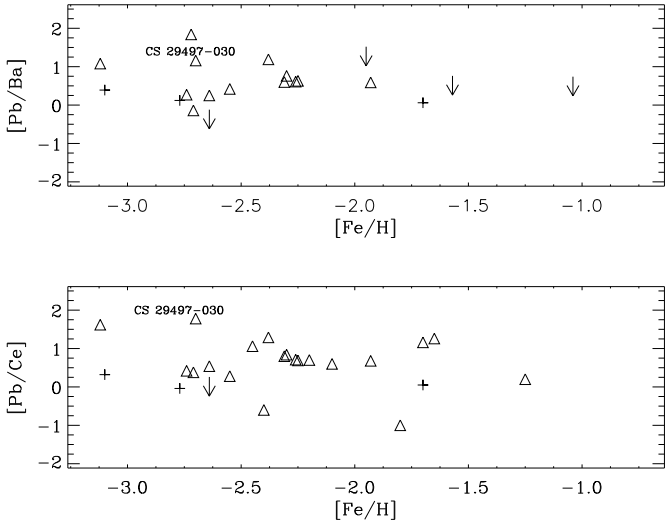}}}
\caption{[Pb/Ba] and [Pb/Ce] ratios as a function
of metallicity.  The data shown are taken from the list in Table 
\ref{pbstars}. The three $r$-process-enhanced stars CS~22892-052, HD~1262238, and
HD~115444, are plotted as crosses. Upper limits are shown as downward arrows.}
\label{pb_ba}
\end{figure}

\begin{figure}
\rotatebox{90}{\resizebox{7.5cm}{!}{\includegraphics{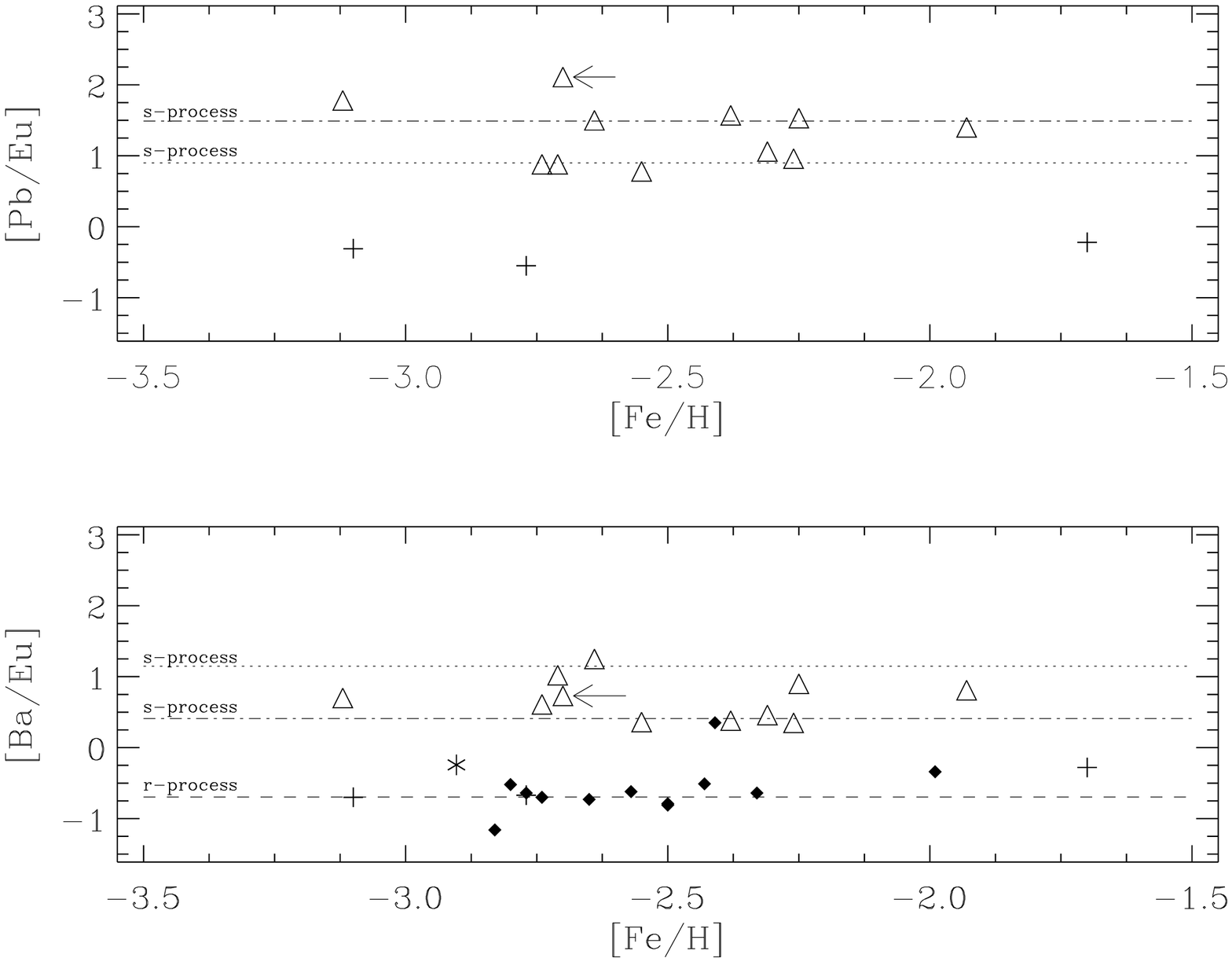}}}
\caption{[Pb/Eu] and [Ba/Eu] ratios,
as a function of metallicity, for the data in Table \ref{pbstars}; the same
symbols are used as in Fig \ref{pb_ba}.  CS~29497-030
is identified by an arrow. Filled symbols are data
from \citet{mcw98}, the asterisk is  CS~31082-001, from \citet{hill}.
The dotted line corresponds to the solar s-process main component, 
while the dashed line corresponds to the inferred solar r-process value \citep{Arlandini}.
The dash-dotted line represents the surface abundance of an AGB model with 
Z=0.001 after 50 thermal pulses \citep{GM2000}.  
}
\label{pbeu}
\end{figure}

\subsection{The chemical history of CS 29497-030 }

The observed abundance pattern of this star strongly suggests the occurrence of
mass transfer in the binary system during the AGB phase of the companion, which
is detected through the present orbital motion of the surviving member (but not
unambigously from the photometry or the spectroscopy). The existing observations
are, however, consistent with the hypothesis that the companion is presently a
white dwarf. The period of about one year implies a rather wide separation,
hence mass transfer must have occurred through a stellar wind (or super-wind) phase 
on the AGB, and is likely to have happened during the thermally pulsating episodes
experienced by the companion.
In this
scenario, it is the third dredge-up that is responsible for bringing to the
surface of the companion the carbon that is freshly produced in the He-burning
shell, as well as the $s-$process elements.

The pattern of neutron-capture elements in CS~29497-030 seems to show
a pure $s-$process signature, as suggested in Fig. \ref{spatt_tab}  which 
compares the abundance pattern in the star with the main $s-$process
pattern from \citet{Arlandini}. All the heavy elements above Z=40 fit 
the pattern rather well, 
except for Pb. 

In Fig. \ref{spatt_tab1} we compare the
abundance pattern of CS~29497-030 to the surface abundances of a model AGB star
with Z=0.001 after 50 thermal pulses, taken from \citet{GM2000}.
The overall agreement for CS~31062-050 is better than in Fig. \ref{spatt_tab}.
For both stars  the large Pb overabundance is reproduced within a factor of 3.
It is, however,  quite striking to note the large underabundance of Sr, Y, and Zr 
compared to both the models and the main solar system s-component.  Several
known s-process-rich, very metal-poor stars have high ratios (more than 1.0 dex) 
of the heavy $s$-process elements (hs: La, Ce, Nd) relative to the lighter $s$-process
elements (ls: Sr, Y, Zr) \citep{hill00}. CS 29497-030 has [Ba/Eu] =+0.73, and
[hs/ls] = +1.02, which is quite similar to other s-process-enhanced, very metal-poor
stars.

In Fig. \ref{gallino} we compare the observed [X/Fe] ratios
in CS~29497-030 and CS~31062-050 with the theoretical
$s-$process computations of Gallino et al. (in preparation).
In the case shown the AGB companion had a Main Sequence mass
of 1.3 M$\odot$. The accretion was modeled as occurring
through stellar wind, rather than Roche-Lobe overflow, 
which seems a reasonable assumption for such a wide system.
The agrement is quite good for the heavy elements, the Zr abundance
is now well reproduced; although the predicted Y and Sr remain too
high. Even more
interesting is that a considerable enhancement of O is predicted
in the model. This arises from the 
$\rm^{12}C(\alpha,\gamma)^{16}O$ reaction occurring in
the He shell. Although the amount of O produced is small,
at such a low metallicity it is sufficient to enhance
the [O/Fe] ratio by about 0.5 dex.

In Table \ref{pbstars} we have summarised the data for all stars with lead
detections. This table includes stars that exhibit predominantly $s$-process
patterns \citep{aoki01,aoki02b,johnson02,lucatello}, and some with
predominantly $r$-process patterns \citep{sneden98,sneden00,travaglio}. Some of
them are known to be binaries, and some are known CH stars \citep{vaneck,
vaneck03}.

In Fig. \ref{pb_ba} we have assembled the reported [Pb/Ba] and [Pb/Ce] ratios
for all the stars listed in Table \ref{pbstars}, which summarizes results for
metal-poor stars where these ratios are available. A large scatter is apparent;
the $r$-process-enhanced stars CS~22892-052, HD~126238, and HD~115444 do not
stand out against the others. In Fig. \ref{pbeu} we show [Pb/Eu] for all of the
metal-poor stars for which both Pb and Eu have been measured. The crosses
represent the three $r$-process-enhanced stars CS~22892-052, HD~1262238, and
HD~115444. It is clear that the $r$-process-enhanced stars stand out from the
others, which follow an $s$-process pattern instead. The r$-$process-enhanced
stars are well below both the 
[Pb/Eu] ratio predicted by the metal-poor AGB models
\citep{GM2000} and the solar $s$-process main component pattern \citep{Arlandini}, 
while all the stars that are close to the horizontal lines also show enhanced
$s$-process elements. CS~29497-030 exhibits a very high [Pb/Eu] ratio (indicated
by an arrow in Fig. \ref{pbeu}). 

In the bottom panel of Fig. \ref{pbeu}, we
show [Ba/Eu] as a function of metallicity for all the stars in Table
\ref{pbstars} for which Ba was measured, as well as for the stars of
\citet{mcw98} (filled symbols) and for CS~31082-001 from \citet{hill}, plotted
as an asterisk. In this figure the $r$-process-enhanced stars stand out with
respect to the $s$-process pattern stars. The comparison with the ``normal''
halo stars of \citet{mcw98}, i.e. those not enhanced in neutron-capture elements,
suggests that these stars follow the
$r-$process pattern. The sole exception is CS~22947-187, a carbon-enhanced,
metal-poor star that falls in the regime occupied by the $s$-process-enhanced
stars. It is interesting to note that CS~31082-001 falls in the middle between the
$s$-process and $r$-process-enhanced stars. 

CS~29497-030 also exhibits a conspicuously large nitrogen overabundance. In terms of
[C/N], however, the star is not exceptional among the Pb stars, which span the range
$\rm -0.1 \le [C/N] \le 1.3$, and in fact has the same [C/N] as HE~0024-2523
\citep{lucatello}.  The possible origin of the nitrogen could be a massive AGB companion,
which can produce N through the Hot Bottom Burning (HBB) process. However, existing 
AGB models predict that the N abundance should be higher than that of C when
HBB is responsible for its production \citep{siess}. Furthermore, the high
$s$-process abundances observed in the Pb-rich stars cannot be obtained by this
scenario. Another possible scenario that can produce C and N is via hot
dredge-up \citep{herwig}. This is a simultaneous action of H-burning and third dredge-up. 
\citet{herwig} discusses models that are able to produce C and N with C $>$ N,
however, the $s$-process elements cannot be produced by this approach. 
\citet{GS01}, by forcing extra mixing and initial pollution of C from the core
He-burning regions, are able to solve the problem.  However, these calculations are
still in the preliminary stage.

The oxygen abundance of CS~29497-030 appears exceptionally
high, both with respect to most other metal-poor stars \citep{israelian01,
depagne,MB02,nissen,barbuy03}, and with respect to HE~0024-2523
\citep{lucatello}. On the one hand it is true that some intermediate-mass stars
may indeed produce sizeable amounts of oxygen \citep{M01}, and perhaps at least
a part of the oxygen we observe has been dumped on the surface of the star by
the same process that contributed carbon and nitrogen. On the other hand, if we
assume the O abundance is {\em intrinsic}, and not {\em photospheric}, then the
total {\em metal} 
abundance $Z$  of CS~29497-030 is not very low, which implies an 
even younger age and further exacerbates the age problem noted above. In this case it is
also difficult to explain why the observed iron abundance is so low.

\begin{table*}
\caption{Summary of available data for lead (Pb) in metal-poor stars}
\label{pbstars}
\begin{tabular}{lllllllllll}
\hline
 Object             & \teff/log g      & [Fe/H]        & [C/Fe]        & [Ba/Fe]     & [Ce/Fe]     &  [Pb/Fe]    & [Eu/Fe] Binary & Period (days)\\
 \hline
 CS~22183-015$^{1}$ & 5200,2.5         & $-$3.12       &  2.2$^2$      &    2.09     & 1.55        & 3.17        & 1.39        &\nodata &\nodata \\ 
 CS~22880-074$^{3}$ & 5850,3.8         & $-$1.93       &  1.3          &    1.31     & 1.22        & 1.9         & 0.5         &    no  &\nodata \\
 CS~22892-052$^{4}$ & 4710,1.5         & $-$3.1        &  1.0$^{5}$    &    0.96     & 1.03        & 1.35        & 1.66        &\nodata &\nodata \\
 CS~22898-027$^{3}$ & 6250,3.7         & $-$2.26       &  2.2          &    2.23     & 2.13        &     2.84    & 1.88        &    no  &\nodata \\
 CS~22942-019$^{3}$ & 5000,2.4         & $-$2.64       &  2.0          &    1.92     & 1.54        &$\le1.6$     & 0.79        &    yes & 2800   \\
 CS~29497-030$^{6}$ & 6650,3.5         & $-$2.70       &  2.38         &    2.17     & 2.14        &     3.55    & 1.44        &    yes & 342    \\
 CS~29526-110$^{3}$ & 6500,3.2         & $-$2.38       &  2.2          &    2.11     & 2.01        &     3.3     & 1.73        &    yes &\nodata \\
 CS~30301-015$^{3}$ & 4750,0.8         & $-$2.64       &  1.6          &    1.45     & 1.16        & 1.7         & 0.2         &\nodata &\nodata \\
 CS~31062-012$^{3}$ & 6250,4.5         & $-$2.55       &  2.1          &    1.98     & 2.12        &     2.4     & 1.62        &\nodata &\nodata \\
 CS~31062-050$^{3}$ & 5600,3.0         & $-$2.31       &  2.0          &    2.30     & 2.10        &     2.9     & 1.84        &\nodata &\nodata \\
 LP~625-44$^{7}$    & 5500,2.8         & $-$2.71       &  2.1          &    2.74     & 2.22        &    2.6      & 1.72        &     yes&\nodata \\
 LP~706-7$^{7}$     & 6000,3.8         & $-$2.74       &  2.15         &    2.01     & 1.86        &  2.28       & 1.40        &     no &\nodata \\
 HD~26 $^{8}$       & 5170,2.2         & $-$1.25       & \nodata       & \nodata     & 1.7         & 1.9         &\nodata      &\nodata &\nodata \\
 HD~2665$^{9}$      & 5061,2.35$^{10}$ & -1.95$^{10}$  & -0.13$^{10}$  &  $-$0.32    & \nodata     & $<1.$       &\nodata      &\nodata &\nodata \\
 HD~21581$^{9}$     & 4978,2.29$^{10}$ & -1.57$^{10}$  & -0.25$^{10}$  &  $-$0.06    & \nodata     & $<0.5$      &\nodata      &\nodata &\nodata \\
 HD~115444$^{11}$   & 4750,1.7$^{11}$  & -2.77$^{11}$  & -0.11$^{12}$  & 0.18$^{12}$ & 0.34$^{12}$ & 0.30$^{11}$ & 0.85$^{12}$ &\nodata &\nodata \\
 HD~126238$^{13}$   & 4979,2.5$^{14}$  & -1.7$^{14}$   &  \nodata      &  $-$0.11    & $-$0.1      & -0.05$^{13}$& 0.17$^{14}$ &\nodata &\nodata \\
 HD~187861$^{15}$   & 5320,2.3         & $-$1.65       &  1.89         & \nodata     & 1.14        &  2.40       &\nodata      &\nodata &\nodata \\
 HD~189711$^{16}$   & 3500,0.5         & $-$1.8        & \nodata       & \nodata     & 1.7         & 0.7         &\nodata      &\nodata &\nodata \\
 HD~196944$^{3}$    & 5250,1.8         & $-$2.25       &  1.2          &    1.07     & 1.01        & 1.7         & 0.17        &\nodata &\nodata \\
 HD~196944$^{7}$    & 5250,1.7         & $-$2.45       &  1.32         & \nodata     & 1.09        &  2.15       &\nodata      &\nodata &\nodata \\
 HD~198269$^{16}$   & 4800,1.3         & $-$2.2        & \nodata       & \nodata     & 1.5         & 2.2         &\nodata      &\nodata &\nodata \\
 HD~201626$^{16}$   & 5190,2.25        & $-$2.1        & \nodata       & \nodata     & 1.8         & 2.4         &\nodata      &\nodata &\nodata \\
 HD~201891$^{9}$    & 5991,4.30$^{10}$ & -1.04$^{10}$  &  0.11$^{10}$  &  $-$0.04    & \nodata     & $<0.5$      &\nodata      &\nodata &\nodata \\
 HD~224959$^{15}$   & 5200,2.3         & $-$1.7        &  1.95         & \nodata     & 1.39        & 2.55        &\nodata      &\nodata &\nodata \\
 HD~23439A$^{9}$    & 5140,4.48$^{10}$ & -0.99$^{10}$  &  0.09$^{10}$  &   0.21      & \nodata     &  0.6        &\nodata      &\nodata &\nodata \\
 HE~0024-2523$^{17}$& 6625,4.3         & $-$2.72       &  2.6          &    1.46     & \nodata     & 3.3         & $<$1.1      &     yes& 3.14   \\
 HE~2148-1247$^{18}$& 6380,3.9         & $-$2.3        &  1.91         &   2.36      & 2.28        & 3.12        &\nodata      &   yes  &\nodata \\
 V~Ari$^{8}$        & 3580,-0.2        & $-$2.4        & \nodata       & \nodata     & 1.6         & 1.0         &\nodata      &\nodata &\nodata \\
\hline
\\

\end{tabular}
\noindent \parbox[t]{15cm}{\small
     (1) \citet{johnson02} 
     (2) \citet{norris97} 
     (3) \citet{aoki02b}
     (4) \citet{sneden00} 
     (5) \citet{barbuy97}
     (6) Present work 
     (7) \citet{aoki01}
     (8) \citet{vaneck03}
     (9) \citet{travaglio}
    (10) \citet{gratton2000}
    (11) \citet{sneden98}
    (12) \citet{westin}
    (13) \citet{cowan}
    (14) \citet{gratton94}
    (15) \citet{vaneck} 
    (16) \citet{vaneck03}
    (17) \citet{lucatello} 
    (18) \citet{cohen03}
  }
\end{table*}

Inspection of Fig. \ref{pb_ba} suggests that the observed [Pb/Ba] ratios 
for metal-poor stars exhibit scatter in excess of that due to
observational errors. It is not clear yet whether this establishes that there 
is a range of operating conditions for the $s-$process at low metallicities.

\section{Conclusions}

The number of lead-rich, extremely metal-poor stars has now grown up to 25.
Collectively, they demonstrate unequivocally that the $s-$process may already
operate even at very low metallicities, and that its occurrence is not  very 
rare. This also suggests that mixing of protons into the C-rich layers in
low metallicity stars does indeed occur, at least in some cases, and that this
has been experienced by the companion of CS~29497-030. Different $^{13}C$
profiles and stellar masses will result in different abundance patterns, as
shown by \citet{busso01}; it is therefore possible that a suitable choice of
both will reproduce the observed heavy element pattern for CS~29497-030 and
other stars like it.

Whether or not these nucleosynthesis products have an impact on the early chemical
evolution of the Galaxy is still an open question. It depends crucially on how many
C- and $s-$process-element producing stars of zero- or extremely 
low metallicity 
existed during the first and second generations of star formation, and on how 
efficiently their products were mixed into the ISM. \citet{burris} provided evidence 
that the $s-$process becomes globally important beginning at [Fe/H] = $\sim -2.7$; 
however, the extremely
metal-poor star CS~22183-015 displays an almost pure $s-$process-element pattern
already at [Fe/H] = --3.12 \citep{johnson02}.  Whether stars such as these are
the exception or the rule at extremely low metallicity will have to be
addressed, based on more complete high-resolution analyses of a larger sample of
stars, in the near future.

The comparison made in Fig. \ref{pbeu} with ``normal'' halo stars suggests that
most of these exhibit elemental abundance patterns that lie close to the
$r-$process pattern. 
More data are needed to draw definitive conclusions; however
the presently available observations suggest that the $r-$only hypothesis of
\citet{truran} as the {\em standard} assumption is questionable, and chemical
evolution models ought to take into account a mixture of $s-$ and $r-$processes 
for the production of the heavy metals at arbitrarily low metallicity.
High-resolution analyses of a larger number of very metal-poor stars will
allow for better constraints on the relative contributions of the two
processes in the early Galaxy.

All of the lead-rich stars with $s$-process enhancement show large carbon
abundances, like the classical CH stars. The fact that many of them appear to be
members of binary systems emphasizes the analogy. A detailed comparison of nitrogen
abundances for these stars could be quite important, in order to set limits on
the likely masses of presumed AGB companions, which could have experienced
different mixing, mass-loss, and nucleosynthesis episodes in the course of their
evolution.

Our analysis has shown that CS~29497-030 is an extreme case of a lead-rich, very
metal-poor star. Its binary nature suggests that the large enhancement of
neutron-capture elements which follow an $s-$process pattern, as well as the
large enhancement of C and N, is in fact due to mass transfer from an
intermediate-mass companion during its AGB phase. The very low iron abundance
implies that $s-$process sources are active at quite early times in Galactic
history. The apparently large O abundance, if real, is not easy to explain, nor
is the overabundance of Na which recalls the similar, but far more dramatic 
enhancement of Na in CS~22949-037 \citep{depagne}. 
The comparison with other lead stars shows a
sizeable scatter in abundance ratios like [Pb/Ba] and [Pb/Ce]. Although the
observational error alone cannot be absolutely excluded as the source of the
observed scatter, the possibility exists that the scatter is real and due to different
physical conditions in the operation of the early $s-$process.
Our preliminary finding of a relatively large axial rotational velocity for 
CS~29397-030, the high T$_{eff}$ of the star, and the lack of detectable Li, suggest
that CS~29397-030 has been ``spun up'' as the result of mass transfer and
should be classified as a halo blue stragger, similar in many respects to the
stars described by \citet{ryan02}.  Detailed analyses of other such stars are 
certain to be illuminating.

\begin{acknowledgements}

We are grateful to A. Korn for performing the
NLTE computations for Fe. 
Special thanks are due to R. Gallino for  illuminating discussions
on $s-$process nucleosynthesis and for providing us the results
of his computations in advance of publication.
We also wish to
acknowledge helpful discussions with H. Schlattl, L. Girardi and S. Zaggia.
Thanks to F. Herwig for his input on the issue of nitrogen.
Finally we thank the referee G. Preston whose comments
contributed to considerably improve this paper.
This research received support from the italian 
MIUR COFIN2002 grant 2002028935\_ 003.
BA and JA thank the Carlsberg Foundation and 
the Swedish and Danish Natural Science
Research Councils for financial support for this work.
T.C.B acknowledges partial support from grants AST 00-98508 and AST 00-98549,
awarded by the U.S. National Science Foundation.
This publication makes use of data products from the 2MASS All Sky Survey,
which is a joint project of the University of Massachusetts and the 
Infrared Processing and Analysis Center/California Institute of Technology,
funded by the National Aeronautics and Space 
Administration and the National Science Foundation.

\end{acknowledgements}

\bibliographystyle{aa}

\begin{table*}
\caption{Line list }
\begin{tabular}{lllllll}
\hline
Element & $\lambda(nm)$ & Low.Exc (eV) & log(gf) & EW(pm) & Ref  \\
\hline
C I  & 437.1367 & 7.68 &-2.333 &  4.33 & 1 \\
C I  & 476.6677 & 7.48 &-2.617 &  1.66 & 1 \\
C I  & 477.0026 & 7.48 &-2.439 &  2.29 & 1 \\
C I  & 477.1746 & 7.49 &-1.866 &  5.50 & 1 \\
C I  & 477.5895 & 7.49 &-2.304 &  2.78 & 1 \\
C I  & 493.2049 & 7.68 &-1.884 &  4.84 & 1 \\
C I  & 502.3839 & 7.95 &-2.209 &  1.63 & 1 \\
C I  & 505.2167 & 7.68 &-1.648 &  7.47 & 1 \\
C I  & 538.0337 & 7.68 &-1.842 &  5.25 & 1 \\
C I  & 579.3124 & 7.95 &-2.062 &  1.12 & 1 \\
C I  & 580.0602 & 7.95 &-2.338 &  1.01 & 1 \\
C I  & 600.1118 & 8.64 &-2.061 &  0.28 & 1 \\
C I  & 601.4834 & 8.64 &-1.585 &  0.97 & 1 \\
C I  & 658.7610 & 8.54 &-1.596 &  3.65 & 1 \\
C I  & 708.7836 & 8.647&-1.443 &  1.81 & 1 \\
C I  & 709.3234 & 8.647&-1.697 &  syn  & 1 \\
C I  & 710.0123 & 8.643&-1.470 &  1.70 & 1 \\
C I  & 710.8930 & 8.640&-1.592 &  1.62 & 1 \\
C I  & 711.1469 & 8.640&-1.086 &  2.96 & 1 \\
C I  & 711.3179 & 8.647&-0.774 &  4.61 & 1 \\
C I  & 711.6988 & 8.647&-0.907 &  4.71 & 1 \\
C I  & 711.9657 & 8.643&-1.149 &  3.36 & 1 \\
O I  & 777.1941 & 9.146& 0.369 &  7.79 & 1 \\
O I  & 777.4161 & 9.146& 0.223 &  6.91 & 1 \\
O I  & 777.5390 & 9.146& 0.001 &  4.18 & 1 \\
Na I & 588.9951 & 0.00 & 0.112 & 10.24 &6\\
Na I & 589.5924 & 0.00 &-0.191 &  7.37 &6\\
Mg I & 416.7271 & 4.346&-1.004 &  syn  &1 \\
Mg I & 517.2684 & 2.71 &-0.380 & 11.64 &6\\
Mg I & 518.3604 & 2.72 &-0.158 & 13.13 &6\\
Mg I & 552.8405 & 4.34 &-0.341 &  2.25 &6\\
Mg II& 448.1126 & 8.864& 0.740 &  syn  &1\\
Mg II& 448.1325 & 8.864& 0.590 &  syn  &1\\
Al I & 394.4006 & 0.00 &-0.640 &  7.49 &1\\
Al I & 396.1520 & 0.010&-0.340 &  syn  &1\\
Si I & 390.5523 & 1.91 &-1.090 &  8.16 &1\\
Si II& 385.6018 & 6.859&-0.557 &  syn  &1 \\
Si II& 386.2595 & 6.858&-0.817 &  syn  &1 \\
S  I & 921.2863 & 6.525& 0.420 &  syn  &1 \\
\hline
\end{tabular}
\end{table*}

\setcounter{table}{5}
\begin{table*}
\renewcommand{\thetable}{\arabic{table} (continued)}
\caption{Linelist}
\begin{tabular}{lllllll}
\hline
Element & $\lambda(nm)$ & Low.Exc (eV) & log(gf) & EW(pm)   & Ref  \\
\hline
Ca I & 422.6728 & 0.00 & 0.240 & 10.79 &6\\
Ca I & 431.8652 & 1.90 &-0.210 &   .94 &6\\
Ca I & 443.4957 & 1.89 & 0.066 &  1.36 &1\\
Ca I & 445.4779 & 1.90 & 0.260 &  3.39 &6\\
Ca I & 558.8749 & 2.52 & 0.210 &  1.23 &6\\
Ca I & 612.2217 & 1.89 &-0.320 &   .91 &6\\
Ca I & 616.2173 & 1.90 &-0.090 &  2.13 &6\\
Ca I & 643.9075 & 2.52 & 0.470 &   .98 &6\\
Ca I & 646.2567 & 2.52 & 0.230 &   .59 &1\\
Sc II& 357.6340 & 0.01 & 0.007 &  5.12 &1\\
Sc II& 361.3829 & 0.02 & 0.416 &  5.95 &1\\
Sc II& 424.6822 & 0.31 & 0.240 &  5.68 &6\\
Sc II& 440.0389 & 0.605&-0.536 &   syn &1\\
Sc II& 441.5557 & 0.595&-0.668 &   syn &1\\
Ti I & 498.1731 & 0.84 & 0.500 &   2.4 &6\\
Ti II& 337.2800 & 0.01 & 0.270 & 12.77 &1\\
Ti II& 338.0279 & 0.05 &-0.570 &  3.71 &1\\
Ti II& 344.4314 & 0.15 &-0.810 &  4.85 &1\\
Ti II& 349.1066 & 0.11 &-1.060 &  3.25 &1\\
Ti II& 350.4896 & 1.89 & 0.180 &  3.66 &1\\
Ti II& 351.0845 & 1.89 & 0.140 &  2.84 &1\\
Ti II& 359.6052 & 0.61 &-1.220 &  3.34 &1\\
Ti II& 391.3468 & 1.12 &-0.410 &  5.44 &1\\
Ti II& 439.5033 & 1.08 &-0.510 &  6.19 &1\\
Ti II& 439.9772 & 1.24 &-1.220 &  2.11 &1\\
Ti II& 441.7719 & 1.16 &-1.230 &  1.61 &1\\
Ti II& 444.3794 & 1.08 &-0.700 &  4.74 &1\\
Ti II& 446.8507 & 1.13 &-0.600 &  4.95 &1\\
Ti II& 450.1273 & 1.12 &-0.760 &  3.89 &1\\
Ti II& 453.3969 & 1.24 &-0.540 &  4.78 &1\\
Ti II& 456.3761 & 1.22 &-0.790 &  4.15 &1\\
Ti II& 457.1968 & 1.57 &-0.230 &  4.75 &1\\
Ti II& 480.5085 & 2.06 &-0.960 &  0.53 &1\\
Ti II& 518.8680 & 1.58 &-1.050 &  1.19 &1\\
Cr I & 357.8684 & 0.00 & 0.409 &  3.75 &1\\
Cr I & 359.3481 & 0.00 & 0.307 &  3.78 &1\\
Cr I & 425.4332 & 0.00 &-0.110 &  2.61 &6\\
Cr I & 427.4796 & 0.00 &-0.230 &  2.70 &6\\
Cr I & 520.6038 & 0.94 & 0.020 &  0.58 &6\\
Cr I & 520.8419 & 0.94 & 0.160 &  1.55 &6\\
Cr II& 335.8491 & 2.45 &-0.722 &  2.76 &1\\
Cr II& 336.8041 & 2.48 &-0.319 &  7.07 &1\\
Cr II& 338.2675 & 2.45 &-0.639 &  2.80 &1\\
Cr II& 340.8757 & 2.48 &-0.038 &  5.27 &1\\
Cr II& 342.1202 & 2.42 &-0.611 &  3.73 &1\\
\hline
\end{tabular}
\end{table*}

\setcounter{table}{5}
\begin{table*}
\renewcommand{\thetable}{\arabic{table} (continued)}
\caption{Linelist}
\begin{tabular}{lllllll}
\hline
Element & $\lambda(nm)$ & Low.Exc (eV) & log(gf) & EW(pm)    & Ref  \\
\hline
Mn I & 403.0753 & 0.00 &-0.480 &  1.58 &6\\
Mn I & 403.3062 & 0.00 &-0.620 &  1.46 &6\\
Mn I & 403.4483 & 0.00 &-0.810 &  2.30 &6\\
Mn II& 344.1988 & 1.78 &-0.270 &  5.18 &1\\
Mn II& 348.8677 & 1.85 &-0.860 &  3.20 &1\\
Fe I & 344.0606 & 0.00 &-0.670 &  6.97 &1\\
Fe I & 344.0989 & 0.05 &-0.960 &  7.47 &1\\
Fe I & 355.4925 & 2.83 & 0.538 &  2.61 &1\\
Fe I & 356.5379 & 0.96 &-0.133 &  5.68 &1\\
Fe I & 360.8859 & 1.01 &-0.100 &  5.86 &1\\
Fe I & 361.0159 & 2.81 & 0.176 &  1.40 &1\\
Fe I & 361.8768 & 0.99 &-0.003 &  5.68 &1\\
Fe I & 362.1461 & 2.73 &-0.016 &  2.03 &1\\
Fe I & 384.9967 & 1.01 &-0.970 &  5.06 &1\\
Fe I & 385.6372 & 0.05 &-1.290 &  7.16 &1\\
Fe I & 385.9911 & 0.00 &-0.710 &  9.88 &1\\
Fe I & 386.5523 & 1.01 &-0.980 &  4.47 &1\\
Fe I & 392.0258 & 0.12 &-1.750 &  4.57 &6\\
Fe I & 400.5242 & 1.56 &-0.610 &  4.67 &6\\
Fe I & 404.5812 & 1.48 & 0.280 &  9.26 &6\\
Fe I & 406.3594 & 1.56 & 0.070 &  7.77 &6\\
Fe I & 407.1738 & 1.61 &-0.020 &  7.39 &6\\
Fe I & 413.2058 & 1.61 &-0.670 &  3.79 &6\\
Fe I & 414.3415 & 3.05 &-0.204 &  1.20 &1\\
Fe I & 414.3868 & 1.56 &-0.460 &  4.48 &6\\
Fe I & 418.1755 & 2.83 &-0.180 &  1.29 &6\\
Fe I & 418.7039 & 2.45 &-0.550 &  1.62 &6\\
Fe I & 418.7795 & 2.42 &-0.550 &  1.91 &6\\
Fe I & 419.1431 & 2.47 &-0.730 &  1.28 &6\\
Fe I & 419.9095 & 3.05 & 0.250 &  2.76 &6\\
Fe I & 420.2029 & 1.48 &-0.700 &  4.50 &6\\
Fe I & 422.7427 & 3.33 & 0.230 &  1.70 &1\\
Fe I & 423.3603 & 2.48 &-0.600 &  1.73 &6\\
Fe I & 425.0119 & 2.47 &-0.400 &  1.65 &6\\
Fe I & 425.0787 & 1.56 &-0.714 &  3.63 &1\\
Fe I & 426.0474 & 2.40 &-0.020 &  4.10 &6\\
Fe I & 427.1154 & 2.45 &-0.350 &  2.69 &6\\
Fe I & 427.1761 & 1.48 &-0.160 &  7.40 &1\\
Fe I & 438.3545 & 1.48 & 0.200 &  8.77 &6\\
Fe I & 440.4750 & 1.56 &-0.140 &  7.08 &6\\
Fe I & 441.5123 & 1.61 &-0.610 &  3.91 &6\\
Fe I & 449.4563 & 2.20 &-1.140 &  0.98 &6\\
\hline 
\end{tabular}
\end{table*}

\setcounter{table}{5}
\begin{table*}
\renewcommand{\thetable}{\arabic{table} (continued)}
\caption{Linelist}
\begin{tabular}{lllllll} 
\hline 
Element & $\lambda(nm)$ & Low.Exc (eV) & log(gf) &  EW(pm)   & Ref  \\
\hline 
Fe I & 473.6773 & 3.21 &-0.750 &  0.40 &1\\
Fe I & 489.1492 & 2.85 &-0.110 &  1.56 &1\\
Fe I & 491.8994 & 2.87 &-0.340 &  1.23 &1\\
Fe I & 492.0503 & 2.83 & 0.070 &  2.76 &1\\
Fe I & 495.7597 & 2.81 & 0.233 &  3.52 &1\\
Fe I & 500.6119 & 2.83 &-0.620 &  0.38 &1\\
Fe I & 501.2068 & 0.86 &-2.642 &  0.14 &1\\
Fe I & 504.9820 & 2.28 &-1.360 &  0.36 &1\\
Fe I & 523.2940 & 2.94 &-0.060 &  1.38 &1\\
Fe I & 526.6555 & 3.00 &-0.390 &  0.42 &1\\
Fe I & 526.9537 & 0.86 &-1.320 &  3.79 &1\\
Fe I & 532.4179 & 3.21 &-0.240 &  1.30 &1\\
Fe I & 532.8039 & 0.92 &-1.470 &  3.13 &1\\
Fe I & 537.1490 & 0.96 &-1.650 &  2.28 &1\\
Fe I & 538.3369 & 4.31 & 0.640 &  0.58 &1\\
Fe I & 539.7128 & 0.92 &-1.990 &  1.28 &1\\
Fe I & 540.5775 & 0.99 &-1.840 &  1.28 &1\\
Fe I & 542.9697 & 0.96 &-1.880 &  1.18 &1\\
Fe I & 543.4524 & 1.01 &-2.120 &  0.93 &1\\
Fe I & 544.6917 & 0.99 &-1.910 &  1.30 &1\\
Fe I & 558.6756 & 3.37 &-0.140 &  0.52 &1\\
Fe I & 561.5644 & 3.33 &-0.140 &  1.05 &1\\
Fe II& 423.3172 & 2.58 &-1.900 &  3.35 &1\\
Fe II& 492.3927 & 2.89 &-1.320 &  4.42 &1\\
Fe II& 501.8440 & 2.89 &-1.220 &  5.43 &1\\
Fe II& 516.9033 & 2.89 &-1.303 &  6.04 &1\\
Fe II& 523.4625 & 3.22 &-2.150 &  0.68 &2\\
Co I & 345.3508 & 0.43 & 0.380 &  2.93 &1\\
Co I & 399.5302 & 0.923&-0.220 &   syn &1\\
Ni I & 338.0566 & 0.42 &-0.170 &  3.71 &1\\
Ni I & 342.3704 & 0.21 &-0.760 &  4.63 &1\\
Ni I & 343.3554 & 0.03 &-0.668 &  3.63 &1\\
Ni I & 347.2542 & 0.11 &-0.810 &  2.90 &1\\
Ni I & 349.2954 & 0.11 &-0.250 &  6.23 &1\\
Ni I & 351.0332 & 0.21 &-0.670 &  3.59 &1\\
Ni I & 351.5049 & 0.11 &-0.211 &  5.04 &1\\
Ni I & 356.6366 & 0.42 &-0.236 &  3.40 &1\\
Ni I & 361.0461 & 0.11 &-1.149 &  1.20 &1\\
Ni I & 361.9386 & 0.42 & 0.035 &  4.25 &1\\
Ni I & 385.8292 & 0.42 &-0.970 &  2.86 &6\\
\hline 
\end{tabular}
\end{table*}

\setcounter{table}{5}
\begin{table*}
\renewcommand{\thetable}{\arabic{table} (continued)}
\caption{Linelist}
\begin{tabular}{lllllll} 
\hline 
Element & $\lambda(nm)$ & Low.Exc (eV) & log(gf) &  EW(pm)   & Ref  \\
\hline 
Sr II& 407.7709 & 0.00 & 0.170 & 11.30 &1 \\
Sr II& 421.5519 & 0.00 &-0.170 &  9.86 &1 \\
Y  II& 360.0741 & 0.180& 0.280 &  syn  &7 \\
Y  II& 360.1919 & 0.104&-0.180 &  syn  &7 \\
Y  II& 361.1044 & 0.130&0.110  &  syn  &7 \\
Y  II& 395.0352 & 0.104&-0.490 &   syn &7 \\
Y  II& 417.7529 & 0.409&-0.160 & syn   &7 \\
Zr II& 339.1982 & 0.16 & 0.463 &  3.85 &1 \\
Zr II& 399.1152 & 0.758&-0.252 &   syn &1 \\
Zr II& 399.8954 & 0.559&-0.387 &   syn &1 \\
Ba II& 413.0645 & 2.722& 0.560 &   syn &3 \\
Ba II& 455.4029 & 0.00 & 0.170 & 16.59 &3 \\
Ba II& 493.4076 & 0.00 &-0.150 & 15.26 &3 \\
Ba II& 585.3668 & 0.60 &-1.010 &  5.34 &3 \\
Ba II& 614.1713 & 0.70 &-0.070 & 10.79 &3 \\
Ba II& 649.6897 & 0.60 &-0.377 &  9.20 &3 \\
La II& 391.6042 & 0.235&-0.485 & syn   & 1 \\
La II& 392.9211 & 0.173&-0.297 & syn   & 1 \\
La II& 394.9102 & 0.403& 0.455 & syn   & 1 \\
La II& 398.8515 & 0.403& 0.138 & syn   & 1 \\
La II& 399.5745 & 0.173&-0.094 & syn   & 1 \\
La II& 404.2901 & 0.927& 0.290 &  syn  & 1 \\
La II& 408.6709 & 0.000&-0.032 &  syn  & 1 \\
La II& 423.8374 & 0.40 &-0.337 &  1.08 & 1 \\
La II& 442.9905 & 0.235&-0.366 &  syn  & 1 \\
La II& 492.0976 & 0.13 &-0.773 &  0.93 & 1 \\
La II& 492.1776 & 0.24 &-0.699 &  1.06 & 1 \\
\hline 
\end{tabular}
\end{table*}

\setcounter{table}{5}
\begin{table*}
\renewcommand{\thetable}{\arabic{table} (continued)}
\caption{Linelist}
\begin{tabular}{lllllll} 
\hline 
Element & $\lambda(nm)$ & Low.Exc (eV) & log(gf) &  EW(pm)   & Ref  \\
\hline 

Ce II& 408.3222 & 0.700& 0.270 & 0.81   & 8 \\
Ce II& 413.3802 & 0.864& 0.370 & 0.89   & 8 \\
Ce II& 413.7645 & 0.516& 0.440 & 1.05   & 8 \\
Ce II& 418.6594 & 0.86 & 0.740 & 2.40   & 8 \\
Nd II& 390.1845 & 0.631&-0.008 & syn   & 1 \\
Nd II& 390.5869 & 0.205&-0.156 & syn   & 1 \\
Nd II& 399.1741 & 0.000&-0.509 & syn   & 1 \\
Nd II& 399.4672 & 0.321&-0.268 & syn   & 1 \\
Nd II& 406.1080 & 0.471& 0.347 & syn   & 1 \\
Nd II& 410.9071 & 0.064&-0.316 & syn   & 1 \\
Nd II& 410.9448 & 0.321& 0.184 & syn   & 1 \\
Nd II& 445.1563 & 0.380& 0.026 & syn   & 1 \\
Eu II& 381.9672 & 0.000& 0.485 & syn   & 4 \\
Eu II& 390.7107 & 0.207& 0.195 & syn   & 4 \\
Eu II& 393.0499 & 0.207& 0.270 & syn   & 4 \\
Eu II& 397.1972 & 0.207& 0.270 & syn   & 4 \\
Eu II& 412.9725 & 0.000& 0.220 & syn   & 4 \\
Eu II& 420.5042 & 0.000& 0.210 & syn   & 4 \\
Pb I & 405.7807 & 1.32 &-0.220 & syn   & 5 \\
Pb I & 368.3462 & 0.969&-0.460 & syn   & 1 \\

\hline
\end{tabular}

\noindent \parbox[t]{15cm}{\small
{\it(1) VALD database (as of June 2002) (2) \citet{biemont}
(3) \citet{McWil95} (4) \citet{lawler} (5) \citet{vaneck03}
(6) Martin et al. (2002) NIST , (7) Hannaford et al. (1981)
(8) D.R.E.A.M. data base www.umh.ac.be/~astro/dream.shtml}}
\end{table*}

\end{document}